\documentclass[11pt]{article}
\usepackage{amsmath,amssymb}
\usepackage{latexsym}
\def\beq{\begin{equation}}
\def\eeq{\end{equation}}
\def\bmul{\begin{multline}}
\def\emul{\end{multline}}

\begin{document}
\title{$SO(5,5)$ duality in M-theory and generalized geometry}
\author{David S. Berman$^\ast$, Hadi Godazgar$^\dagger$  
and  Malcolm J. Perry$^{\dagger}$ \\\
\\ $\ast$ Department of Physics,\\Queen Mary\\Mile End Road\\
London E4 9NS,\\England\\ \\
$\dagger$ Department of Applied Mathematics and Theoretical Physics,\\
Centre for Mathematical Sciences,\\
University of Cambridge,\\
Wilberforce Road,\\
Cambridge CB3 0WA,\\ England.}

\maketitle
\begin{abstract}
We attempt to reformulate eleven dimensional supergravity
in terms of an object that unifies the three-form and the metric and makes the M-theory duality group manifest. This short note deals with the 
case of where the U-duality group $SO(5,5)$ acts in five spatial dimensions.

\end{abstract}

\newpage

\section{Introduction}

In \cite{davidmalcolm}, a generalized metric was introduced from
considering duality from the perspective of 
the membrane world volume theory. 
This naturally combined the metric and the three form of
eleven dimensional supergravity into one geometric object.
The space on which the metric acts will be the usual spacetime extended by $n$ dimensions whose coordinates are $p$-forms. The details of 
how many dimensions one needs to add ie. $n$ and which $p$-forms to use for the coordinates is dependent on several details that we describe in the next section.

This metric is a generalization of the
generalized geometry introduced by Hitchin et al \cite{Hitchin} 
and developed in M-theory by Hull \cite{Hull} and Pacheco and Waldram
\cite{Waldram}. 
A key goal of the work of \cite{davidmalcolm} was to encode the
equations of motion of the space on which the duality group acts 
(when there is no dimensional reduction) in terms of the generalized
metric. This then makes the duality group a manifest global symmetry of the action.
Note, we are making no assumptions about the space such as the existence of Killing Vectors.
 
In \cite{davidmalcolm} the duality group was taken to be $SL(5)$ 
corresponding to the U-duality group of six dimensions. 
The four dimensional space on which this duality group acts was 
extended to ten dimensions with the additional six {\it{dual}} 
dimensions having coordinates described by two forms on the 
four dimensional space.

In this note, we will extend this to five spatial dimensions where the
duality group is $SO(5,5)$. 
In doing so we will have to produce the appropriate generalized
metric that combines $g_{ij}$ and $C_{ijk}$ and determine the action that will reduce to the standard
action of eleven dimensional 
supergravity.

The reduction is when we make the {\it{section condition}} that the fields are independent of {\it{dual}} coordinates.
This is a necessary condition for consistency but it is expected that a more general notion of the {\it{section condition}}
can be constructed and different solutions will correspond to different duality equivalent theories.

Let us examine the analogy with Kaluza Klein theory to describe these ideas. 
Starting with Einstein Maxwell theory in four dimensions with four 
dimensional diffeomorphism symmetry and U(1) gauge symmetry one asks whether this can be reproduced by a higher dimensional 
theory of gravity alone with diffeomorphism symmetry in higher dimensional 
space. The answer is of course yes. One removes dependence of the 
metric on the fifth dimension and the theory reduces to Einstein 
Maxwell with the vector field of electromagnetism being the off diagonal 
component of the metric and its gauge symmetry a remnant of the 
five dimensional diffeomorphism. Once this has been constructed 
it is then tempting to remove the Kaluza Klein reduction and 
consider the full five dimensional theory.

Now we have in eleven dimensional supergravity with a metric and 
a 3-form abelian gauge potential, $C_{3}$. Can one play the same 
trick and find an extended space in which a reduction will 
produce our original theory. No ordinary Kaluza Klein theory can do this. 
Yet if we construct the extended space to have two 
form coordinates as one does in generalized geometry then 
one can write an action on the extended space that will 
reduce a la Kaluza and Klein to the usual eleven dimensional 
supergravity where the C-field is again the off diagonal 
component of the metric and its gauge transformation a 
remnant of the diffeomorphisms on the extended space.

Other work along these lines has been done recently by Hohm, Hull and Zwiebach\cite{hull2} for generalized geometry in string theory and by Hillmann
\cite{hillmann} in M-theory for the group $E_7$ and by West \cite{west} for the relationship with M-theory to the IIA string. The general set up of considering the role of the duality groups in 11-dimensional supergravity has been discussed over a period of years by amongst others \cite{Julia:1980gr,TM:1980,NdeW,msymm}. Recent work studying the idea of curvatures in generalized geometry has appeared in \cite{MIT,Korea} and for the Heterotic string in \cite{hetstring}.

\section{Generalized Metrics and their dynamics}

Generalized metrics are metrics on an extended tangent space.
In the initial work of Hitchin corresponding to string theory the tangent space was extended as follows:
\beq
T\Lambda^1(M)  \rightarrow T\Lambda^1(M) \oplus T\Lambda^{*1}(M)
\eeq
where the addition of the cotangent space $T\Lambda^{*1}(M)$ can be viewed as adding in the space of string windings.
In M-theory, the appropriate extension should correspond to windings
of membranes and where possible fivebranes.\footnote{Also for 7
  dimensions and above six-brane windings are relevant.} 
Thus, since we wish to consider the duality group corresponding to
five dimensions 
we must extend the tangent space to include fivebrane modes as follows:
\beq
T\Lambda^1(M)  \rightarrow T\Lambda^1(M) \oplus T\Lambda^{*1}(M) \oplus T\Lambda^{*5}(M) \, .
\eeq
The dimension of the extended space is now sixteen dimensional and so we
seek a metric on this sixteen dimensional space. 
We know that the theory possesses $SO(5,5)$ duality symmetry so we are
left with the group theoretic problem of finding an object that acts on the {\bf{16}} dimensional representation of SO(5,5).
Note, in string theory the T-duality symmetry is $O(d,d)$ and 
one might imagine that this is therefore the same; however, in that
case the representation is 10 dimensional and the generalized metric acts on the {\bf{10}} of SO(5,5). Here in M-theory we act on the {\bf{16}}.

In a future work we will detail the general
construction of how one determines this metric for all the
relevant groups. For now we simply state the result as coming from a group theoretic construction with the above considerations.

The generalized metric is (upper case Latin indices run from 1 to 16):
\beq
M_{IJ} = \begin{pmatrix} g_{ab}+\frac{1}{2} C_a{}^{ef} C_{bef} + \frac{1}{16}    X_a X_b  &
\frac{1}{\sqrt{2}} C_a{}^{mn} + \frac{1}{4 \sqrt{2} } X_a V^{mn} & \frac{1}{4} {g}^{-{1 \over 2}} X_a \\ \frac{1}{\sqrt{2}} C^{kl}{}_b + \frac{1}{4 \sqrt{2} }  V^{kl} X_b  & g^{kl,mn}+ \frac{1}{2} V^{kl}V^{mn} & \frac{1}{\sqrt{2}}  g^{-{1 \over 2}} V^{kl} \\
\frac{1}{4} g^{-{1 \over 2}} X_b & \frac{1}{\sqrt{2}} g^{-{1 \over 2}} V^{mn} & g^{-1} 
\end{pmatrix} \, ,
\label{genmet}
\eeq
where $g_{ab}$ is the usual metric on the five dimensional space (the lower case latin indices run from 1 to 5); $C_{abc}$ is the three form potential of eleven dimensional supergravity, and $ g^{mn,kl}=\frac{1}{2}(g^{mk}g^{nl}-g^{ml}g^{nk})$ which may be used to raise an antisymmetric pair of indices. We also define: 
\beq
V^{ab}=\frac{1}{6} \epsilon^{abcde} C_{cde} \, ,
\eeq
with $\epsilon^{abcde}$ being the totally antisymmetric tensor and
\beq
X_a= V^{de} C_{dea} \, .
\eeq
Note, that the metric (\ref{genmet}) is quartic in $C_{3}$, and contains terms involving the metric and also terms second order in the inverse metric. It acts on a sixteen dimensional space of which the first 5 dimensions are the usual space dimensions.
 
We now attempt to reconstruct the dynamical 
theory out of this generalized metric.
Consider the following Lagrangian,
\begin{multline}
L = g^{1/2} \Biggl(\frac{1}{16} M^{MN} (\partial_M M^{KL})( \partial_N
M_{KL} ) - {\frac{1}{2}} M^{MN} (\partial_N M^{KL}) (\partial_L M_{MK}) \\
+\frac{3}{128}   M^{MN} (M^{KL} \partial_M M_{KL})(M^{RS} \partial_N M_{RS}) 
-\frac{1}{8}    M^{MN} M^{PQ}(M^{RS} \partial_P M_{RS}) (\partial_M M_{NQ}) 
\Biggr)\,  \label{lagrange}
\end{multline}
where $\partial_M = \bigl(\frac{\partial}{\partial x^a},
\frac{\partial}{\partial y_{ab}}, \frac{\partial}{\partial z} \bigr)$.

$L$ can then be evaluated using the definition of the generalized
metric $M_{MN}$, described above, in terms of the usual metric and three form.

We have extended our space to allow the encoding of the duality group
as a geometric symmetry of the system. 
Obviously we must restrict the physical dimension of the space back
down to five dimensions. 
Thus we need to supplement the action with a physical {\it{section condition}}.

This {\it{section condition}} should be a group covariant. 
One natural choice of condition is simply that all fields 
are independent of the additional coordinates $\{ y_{ab} \}, z $, ie. on all fields
\beq
\frac{\partial}{\partial y_{ab}}=0 \, , \qquad  \frac{\partial}{\partial z}=0 \,.
\label{sc}
\eeq

This is not a covariant condition but it is a sufficient condition. 
Restricting the fields in this way should then lead to the action reproducing the usual action for the metric and C-field.

A challenge is to determine the $SO(5,5)$ section condition in
generality. 
Note, that for the string this is known; there one simply uses the
metric of $SO(5,5)$ to construct a projector that halves the dimension
of the space. Here we need to go from sixteen to five dimensions and a simple projection equation of that sort won't suffice.

We then evaluate L in terms of $g_{ab}$ and $C_{abc}$ with the section condition (\ref{sc}). 
After a long and careful calculation, the result, up to a total derivative, is
\beq
L = g^{1/2}(R(\gamma) - {\frac{1}{48}} F^2) \, .  
\eeq
Hence we recover the usual Lagrangian for (the Bosonic sector of) supergravity. 

The Lagrangian (\ref{lagrange}) is a Lagrangian only for the
directions in which we allow a duality transformation to act. 
For the other directions, the kinetic terms maybe written as follows.
\beq
T= g^{1/2}(-\frac{1}{16}({\it tr}\ \dot M^{-1} \dot M)
-\frac{3}{128}({\it tr}\ M^{-1}\dot M)^2) \label{eq:ke} \, .
\eeq

This paper shows that the generalized metrics can be used to formulate the dynamics of the theory. This is an extension of the Kaluza Klein theory where the additional dimensions have $p$-form coordinates and we have rewritten gravity with a 3-form gauge field in terms of  the dynamics of a single generalized metric $M_{IJ}$ acting on the extended space. At this stage, this is simply a rewriting of supergravity. The suggestive thing about Kaluza Klein theory was that once Einstein Maxwell theory was rewritten as a five dimensional theory it inspired the study of higher dimensional theories where the reduction condition was dropped. This is unlikely to make sense for M-theory. In the string theory context the reduction occurs as a particular solution to the section condition which has its origins in the level matching of the closed string theory. What the physical section constraint is in general for M-theory is the most important question for this approach.

\section{Acknowledgements}

We wish to thank Peter West for interesting discussions.
DSB wishes to thank DAMTP, University of Cambridge for continued 
hospitality and is supported
in part by the Queen Mary STFC rolling grant ST/G000565/1. HG is supported by an STFC grant. MJP is in part supported by an STFC rolling grant.
We would like to thank the Mitchell family for their generous hospitality
at Cook's Branch Nature conservancy where some of this work was carried out.
MJP would like to thank Mitchell foundation and Trinity College Cambridge 
for their support.

\end{document}